\documentclass[preprint2]{aastex7}
\usepackage{amsmath}
\usepackage{placeins}
\usepackage{gensymb}

\begin{document}

\title{Tidal evolution of packed moon systems around an Earth-mass planet}

\author[orcid=0009-0005-1722-1971,gname=Alan,sname=Briseno]{Alan Briseno}
\affiliation{Department of Physics and Astronomy, East Texas A\&M University, Commerce, TX 75428, USA}
\email{Abriseno1@leomail.tamuc.edu}  

\author[orcid=0000-0002-9644-8330,gname=Billy,sname=Quarles]{Billy Quarles}
\affiliation{Department of Physics and Astronomy, East Texas A\&M University, Commerce, TX 75428, USA}
\email{billylquarles@gmail.com}

\author[orcid=0000-0003-0216-559X,gname=Marialis,sname=Rosario-Franco]{Marialis Rosario-Franco}
\affiliation{Department of Astrophysical and Planetary Sciences, University of Colorado Boulder, Boulder, CO 80309, USA}
\email{marialis.rosariofranco@colorado.edu}

\begin{abstract}
While missions have long targeted terrestrial exoplanets within the habitable zone of their host stars, the number of exomoon candidates is expected to grow as next generation space-based observatories achieve the photometric sensitivity required to detect their transit signals. Constraining the stability limits of tightly packed moon systems is therefore essential for transit searches and predicting the number of moons around terrestrial planets.  In our Solar System, only three moons orbit the terrestrial planets, motivating the question of whether Earth-mass exoplanet systems can sustain long-lived, tightly packed satellites. We investigate the stability limits of an Earth-mass planet orbiting a Sun-mass star, where the planet hosts multiple moons. We use the \texttt{REBOUND} N-body integrator along with the \texttt{tides\_spin} module in \texttt{REBOUNDx} to assess the stability of tightly packed systems of Luna-, Pluto-, and Ceres-mass moons across a range of tidal dissipation parameters, up to $10^{7}$ dynamical orbits of the innermost moon. We find that an Earth-mass planet can stably host up to two Luna-mass moons, three Pluto-mass moons, or five Ceres-mass moons. Under Earth-like dissipation, the Luna, Pluto, and Ceres packed systems survive within narrow regions of orbital spacing. These results imply that long-lived multi-moon systems around Earth-mass planets are possible but strongly depend on tidal dissipation; for these architectures to exist on billion-year timescales, tides must be weaker than those of the present-day Earth-Moon system.
\end{abstract}


\section{Introduction}

Numerous moons \citep{JPL_SSD_2026} are known to orbit planets and dwarf planets in the Solar System. Most orbit gas giants in the outer Solar System (e.g., Jupiter and Saturn), while only three orbit terrestrial planets: Earth (Luna) and Mars (Phobos and Deimos), suggesting that giant-planet and terrestrial planet moon systems may follow different formation pathways and orbital evolution histories. Moons around gas giants are likely to form via accretion in a circumplanetary disk \citep{Canup2006,Batygin2020}{, which is fed by gas flowing in from the surrounding protoplanetary disk \citep{Ward2010}. Under typical nebular conditions, a protoplanet approaching Saturn mass begins to open an appreciable gap in the protoplanetary disk \citep{Armitage2011}. Gap opening alters the inflow and shear surrounding the planet, allowing a distinct, rotationally supported circumplanetary disk to develop as the gas cools. The resulting disk provides the structure and thermodynamic conditions required for solids to concentrate and accrete into multiple regular satellites.} Irregular satellites may instead be captured into a planet's gravitational field \citep{Pollack1979,Jewitt2005,Nesvorny2007}, producing irregular orbits (e.g., large semimajor axes, high eccentricities, and varied inclinations).{ Our Solar System has shown that multi-moon systems surrounding giant planets are common.}

{However, moon formation around terrestrial planets are likely driven by impacts, ranging from numerous smaller impacts to a single giant impact \citep{Canup1998,Canup2004}.  In particular, the leading explanation for the origin of the Earth--Moon system involves a giant impact that placed a substantial amount of debris into orbit around the proto-Earth \citep{Ida1997,Canup2001,Canup2004,Canup2012}. Although most previous models were designed to reproduce a final system containing one Moon, the intermediate impact-generated disk need not initially accrete into only one satellite \citep{Canup1999}.  Material spreading beyond the fluid Roche limit can accrete sequentially into multiple moonlets, and interactions among those moonlets subsequently determine whether they merge, reimpact the planet, or survive as separate satellites \citep{Canup1999,Crida2012}. In addition, successive impacts can each produce an orbiting debris disk and moonlet, allowing multiple moonlets to coexist while undergoing outward tidal migration \citep{Rufu2017}.  Impact-generated, fractionally large moons are expected to form most readily around rocky planets with $M_p \leq 6\ M_\oplus$ or $R_p \leq 1.6\ R_\oplus$ \citep{Nakajima2022}. An Earth-mass planet therefore lies within the planetary regime in which large impact-generated satellites can form.}

{These mechanisms provide two possible pathways for assembling a packed two-moon system around an Earth-mass planet. A sufficiently massive impact-generated disk could accrete into two large moons rather than one, or two successive giant impacts could each produce a large moonlet \citep{Canup1999,Rufu2017}. If at least two such objects attain approximately Luna mass and avoid immediate merger or reaccretion onto the planet, they could form an initial two-moon architecture. We do not model the formation probability of this outcome; instead, we investigate whether multiple moon systems, once assembled, can survive their subsequent gravitational and tidal evolution.}

Over the past three decades, the discovery of thousands of exoplanets has substantially increased interest in whether potential moons orbit these worlds. Although moons are relatively common in the Solar System, no exomoon has yet been definitively confirmed, despite two promising candidates identified through transit observations, including Kepler-1625b-i and Kepler-1708-i \citep{Teachey2018,Kipping2022}. As this candidate sample grows, theoretical work on exomoon formation, orbital stability, and potential habitability becomes essential to constrain the dynamical limits to which these satellites can actually exist \citep{Domingos2006,Heller2014,Rosario-Franco2020,Kral2026}.

Systems of moons exist orbiting the gas giants, where similar moon systems (in terms of the orbital architecture) could be orbiting terrestrial exoplanets. Planet and moon formation is a chaotic process, where the deficit of moons in our Solar System is likely due to chance events during formation and is further sculpted by dynamical evolution over time. Therefore, determining the dynamical stability of packed moon systems is essential to assess the diversity of satellite architectures that may exist \citep{Heller2014,Rosario-Franco2020}.

In compact systems, small changes in semimajor axis can shift moons into and out of mean-motion resonances (MMRs), which can rapidly excite eccentricities and increase the likelihood of close encounters. The survivability of packed moon systems, therefore depends not only on their initial formation, but also on evolutionary processes that drive outward orbital migration. Key stability boundaries have been established through a combination of analytical and numerical methods. Specifically, \citet{Gladman1993} defined the analytical limits for the Hill stability of two-planet systems, while \citet{Chambers1996} extended this analysis to multi-planet systems, demonstrating that the time to instability scales exponentially with the initial mutual Hill spacing ($\beta$). Subsequent studies have further refined these limits by considering the onset of resonance overlap in tightly packed equal-mass moon systems \citep[e.g.,][]{Deck2013,Wisdom1980,Obertas2017}.

For packed moons, \citet{Satyal2022} found that an Earth-mass planet orbiting a Sun-mass star can theoretically host up to $3 \pm 1$ Luna-mass moons, $4 \pm 1$ Pluto-mass moons, or $7 \pm 1$ Ceres-mass moons. However, those packed moon systems were evaluated using only gravitational forces (i.e., pure N-body integrations), whereas we include both gravitational and tidal forces. Coupled perturbations of both stellar and planetary tidal forces are likely to impact the dynamical evolution of satellite systems. In the Earth-Moon system, tidal dissipation transfers angular momentum from the planet's rotation to the lunar orbit, causing the Moon to migrate outward over time \citep{Murray1999,Barnes2017,Lainey2020}. In multi-moon systems, this outward migration can occur at different rates for each satellite, leading to convergent or divergent orbital evolution \citep{Crida2012}. This process can create a "path to instability" \citep{Petit2020}: as moons are driven through MMRs, gravitational interactions can excite eccentricities and lead to scattering, collisions, or ejection \citep{Wisdom1980,Mudryk2006}.

{In studies of multi-planet systems, machine learning (ML) models are widely utilized to bypass the computational cost of traditional N-body simulations, allowing for rapid dynamical studies of planetary systems.} While modern {ML} software like SPOCK \citep{Tamayo2020b} can estimate the stability for 3-planet systems, they are primarily trained on systems with a central star rather than a planetary host and are not yet optimized for compact systems with tidal forces. Thus, understanding how tidal forces affect the architecture of packed satellite systems is essential for determining whether multi-moon systems around terrestrial planets can survive over astronomical timescales.

In this work, we classify moon-packing using the mutual-Hill spacing parameter $\beta$ and explore stability as a function of $\beta$ and the tidal time lag $\tau${, which controls the tidal dissipation rate and therefore the orbital evolution rate.} We investigate the stability limits of an Earth-mass planet hosting multiple moons when both N-body interactions and tidal forces are included. Our simulations use the \texttt{REBOUND} N-body integrator with the \texttt{tides\_spin} module in \texttt{REBOUNDx} to assess the stability of tightly packed systems of Luna-, Pluto-, and Ceres-mass moons across a range of tidal dissipation parameters up to $10^{7}$ orbits of the innermost moon.

This paper investigates the long-term stability of packed moon systems through a series of numerical simulations. To determine how these systems evolve, Section~\ref{sec:meth} establishes our numerical setup, initial and stopping conditions, and tidal evolution parameters. In Section~\ref{sec: Results}, we present the results of our numerical simulations. We conclude in Section~\ref{sec: Conclusions} by summarizing our findings and discussing their implications on future work \& observations. 

\section{Methods} \label{sec:meth}
\subsection{Numerical Simulation Setup}

We use the N-body software \texttt{REBOUND} \citep[][ver. 4.4.6]{Rein2012} with the extension \texttt{REBOUNDx} \citep[][ver. 4.4.1]{Tamayo2020} to examine the dynamical stability for systems of equal-mass moons around an Earth-mass planet ($M_{\oplus} = 3.003 \times 10^{-6}\ M_\odot$) that orbits a Sun-mass star. In addition to its mass, the planet adopts Earth-based parameters, including Earth's radius, density, moment of inertia, obliquity, and Love number, where we assume a thin atmosphere exists that has a negligible tidal effect. The planet begins on a mildly eccentric ($e=0.0167$) orbit with a semimajor axis of $1\, {\rm au}$, where the perturbations from the host star induce a small forced eccentricity on the surrounding moons \citep{Andrade-Ines2017, Quarles2021}. We employ the IAS15 integrator \citep{Rein2015} within \texttt{REBOUNDx}, which provides adaptive step-size control for modeling the dynamical evolution of planets and their moons orbiting a host star. IAS15 is proven efficient and reliable \citep{Holman2023}, especially for systems involving close encounters. Although IAS15 uses an adaptive timestep, we set the initial timestep for each simulation to 5\% of the orbital period at $1.8\ R_{\rm Roche}$, to maintain consistency with prior work, as well as ensuring adequate resolution of the shortest dynamical timescale in the system.

Each simulation is evaluated up to $10^{7}\ P_1$, which is consistent with previous studies of tightly packed systems \citep[e.g.,][]{Chambers1996,Obertas2017,Satyal2022}. The moons experience tidal migration, interactions with other bodies, and MMRs, where each effect occurs on timescales far exceeding the mean-motion of the moons themselves \citep{Chambers1996,Obertas2017}. We check the system's fate to determine whether the system state has reached instability, where we record the spacing parameter $\beta$, the maximum eccentricity reached by each moon, and the system’s lifetime.

\subsection{Stopping Conditions} \label{sec: Stopping Conditions}
We stop a simulation and record a system outcome if any of the following conditions for a moon are reached: (i) it undergoes a close encounter within the planet's Roche radius of the planet ($d < R_{\rm Roche}$) resulting in tidal disruption, (ii) it attains an escape orbit ($e\geq1$), or (iii) it reaches a semimajor axis beyond 0.4 times the planetary Hill radius, where the host star is efficient at removing (via scattering or ejection) a moon \citep{Domingos2006,Rosario-Franco2020}. Packed moon systems are sensitive to resonances, and even small excitations in eccentricity can quickly cause instability. These conditions allow us to determine how and when the system becomes dynamically unstable. 

The Roche limit sets a boundary on how close a satellite can orbit its host planet through tidal forces. Once a satellite orbits interior to the Roche limit, these tidal forces overwhelm the satellite's self gravity, which can tear it apart. In our simulations, we use the fluid definition of the Roche limit $R_{\rm Roche}$ given by
\begin{align}\label{Roche_eqn}
{R_{\rm Roche} = 2.44 R_p\left(\frac{\rho_p}{\rho_{\rm sat}}\right)^{1/3} },
\end{align} 

\noindent where $R_{p} = 1\ R_{\oplus}$ (i.e., $1$ Earth radius) and $\rho_{p} =  \rho_{\oplus}$ are the radius and density of the planet and $\rho_{\rm sat}$ is the density of the moon. We assume an Earth-like density ($\rho_\oplus = 5.515\, {\rm g/cm^3}$), where the assumed properties for each moon-type are given in Table \ref{tab:moon_params}.

\begin{deluxetable}{cccc}
\tabletypesize{\small}
\tablecolumns{4}
\setlength{\tabcolsep}{8pt}
\tablewidth{0pt}
\renewcommand{\arraystretch}{1.2} 
\tablecaption{ Assumed Properties for Each Moon Category \label{tab:moon_params}}
\tablehead{\colhead{Moon}  & \colhead{Mass ($M_\oplus$)} & \colhead{Radius ($\rm km$)} & \colhead{Density $\rm (g/cm^3$)}  }
\startdata
Luna & 0.0123 & 1737 & 3.3
\\\hline
Pluto & 0.0022 & 1190 & 1.86
\\\hline
Ceres & 0.00015 & 473 & 2.2
\enddata
\tablecomments{Initial satellite parameters (Mass, Radius, and Density) used in all systematic simulations}
\end{deluxetable}

An outer stability limit is set by the Hill radius ($R_{\rm H}$), a region where the planet's gravity dominates over the host star \citep{Hill1878}, which is given as

\begin{align}\label{2}
{R_{\rm H}=a_p\left(\frac{M_p}{3M_\star}\right)^{1/3} },
\end{align}

\noindent where $a_p$ is the planet's semimajor axis and $M_p$ is the planet's mass. For our Earth-Sun system, we use Earth-like parameters (i.e., $M_p = 1\ M_\oplus$ and $a_p = 1\, {\rm au}$) and a Sun-mass star, $M_{\star} = 1\ M_\odot$. Prior studies show that the stability limit for single moons is roughly $0.4-0.5\,R_{\rm H}$ \citep{Domingos2006,Rosario-Franco2020}.  The more conservative limit ($a_c \approx 0.4 R_{\rm H}$) was obtained using a wider range of starting conditions than the canonical limit ($a_c = 0.5R_{\rm H}$). As a result, we use the conservative limit because the outward migration of the outermost moon will effectively present a random starting phase when it enters the chaotic zone.

\subsection{Initial Conditions}

We begin the innermost moon at $a_{1} = 2R_{\rm Roche}$, {a convenient near-inner-edge location motivated by impact-generated disk models in which satellite material accretes after spreading beyond the Roche limit \citep{Ida1997,Salmon2012}. We do not assume that $2R_{\rm Roche}$ is a unique predicted formation radius; rather, it places the moon safely outside the tidal-disruption boundary while remaining near the expected initial accretion region.} The migration time between $1R_{\rm Roche}$ and $2R_{\rm Roche}$ is negligible compared with our integration timescale. Each moon begins on a circular, coplanar{, prograde} orbit around the Earth-mass planet. The system is systematically arranged using the mutual Hill radius. This procedure has been used for dynamical studies of packed systems \citep[e.g.,][]{Chambers1996,Smith2009,Satyal2022}, where we apply a formalism used by \cite{Satyal2022}. Each successive moon semimajor axis $a_{j+1}$ is determined using the relation

\begin{align}\label{4} 
{a_{j+1}=a_{j}\left(\frac{1+\beta X}{1-\beta X}\right)},
\end{align}

\noindent where $\beta$ is a dimensionless spacing parameter, $a_j$ is the semimajor axis of the previous moon, and $X$ is a mass scaling parameter that depends on the mass of assumed moon-type given by

\begin{align}\label{3}
X = \frac{1}{2}\left[\frac{M_1+M_2}{3(M_{\oplus}+jM_{\rm sat})}\right]^{1/3},
\end{align}

\noindent where we calculate starting $j = 1$ to find the semimajor axis $a_2$ for the next moon. The formula is recursive, so that the semimajor axis of the $j+1$th moon depends on the previously determined value. Since all moons in a given simulation have equal mass, we replace $M_{1} + M_{2}$ with $2M_{\rm sat}$. The spacing parameter $\beta$ has been widely used in previous studies of planet packing around individual stars \citep{Smith2009,Chambers1996,Obertas2017} and binary systems \citep{Quarles2018}. The spacing parameter normalizes the separation between two nearby orbits while using the mutual Hill radius described in Eqn. \eqref{4}.

\cite{Gladman1993} shows a minimum spacing of $\beta_{\rm min} = 2\sqrt{3}$ for two equal-mass bodies orbiting a much more massive central body. Smaller values of $\beta$ are unstable due to the overlap of first-order mean-motion resonances (MMRs; \citet{Wisdom1980,Mudryk2006}). \citet{Quarles2018} developed a formula to estimate the maximum spacing for a packed system of S-type planets in a binary system.  We have adapted the formula for our similar architecture through the following,

\begin{align}\label{5}
\beta_{\rm max} = \left(\frac{(a_N/a_1)^{\frac{1}{N-1}} - 1}{(a_N/a_1)^{\frac{1}{N-1}} + 1}\right) \left(\frac{12M_{\oplus}}{M_{\rm sat}}\right)^{1/3} ,
\end{align}

\noindent which depends on the number of moons $N$, the innermost orbit $a_{1}$ and the outermost orbit $a_{N}$. We use the stability limit $a_c = 0.4\ R_{\rm H}$ for $a_N$ to determine $\beta_{\rm max}$ \citep[i.e.,][]{Satyal2022}. We vary $\beta$ from $2\sqrt{3}$ to $\beta_{\rm max}$ in steps of $0.01$. Although this step size may hinder very narrow regions of stability, such regions are dynamically sensitive. The timescale of tidal migration is relatively quick over the course of the simulation, meaning any moon orbiting within a narrow stability region would more than likely migrate into an unstable configuration.

The initial orbital phase of each moon starts with

\begin{align} 
{M_{j} = 2\pi j\varphi\mod 2\pi},
\end{align}

\noindent where $j$ is an integer starting at $1$ following \citet{Smith2009,Quarles2018}. Since no moon pair shares the same initial orbital phase, this reduces the likelihood of biasing towards resonance locking or clustering early within a given simulation, especially when using irrational numbers. Conveniently, the most irrational number is the golden ratio $\varphi = (1+\sqrt{5})/2$. 

To constrain the initial number of moons, we follow \citet{Satyal2022}, where they found an Earth-mass planet can host up to $7 \pm 1$ Ceres-mass, $4 \pm 1$ Pluto-mass, and $3 \pm 1$ Luna-mass moons without tides. Therefore, we examine systems with five Ceres-mass moons, three Pluto-mass moons, and two Luna-mass moons (see Table \ref{tab:moon_params} for assumed moon properties).

\subsection{Tidal evolution}

Tidal forces arise from differences in gravitational force and scale as $F_{tidal}\propto M/d^{3}$. Both stellar and lunar forces produce tidal bulges that mediate the transfer of angular momentum between the planet and its moons. Consequently, in a packed moon system, tidal forces can {compromise long-term stability by causing moons to migrate into unstable regions/resonances.} Outward tidal migration represents a change in the semimajor axis of each moon, where the inner satellites evolve faster than the outer ones.

We use the \texttt{REBOUNDx} \texttt{tides\_spin} module \citep{Lu2023}, which implements a constant time-lag (CTL) tidal model \citep{Hut1981,Eggleton1998,Leconte2010}. Tidal dissipation is parameterized by the time lag, $\tau$, which represents the lag between the tidal potential and the tidal bulge. To use this module, we must specify a constant tidal time lag $\tau$, the normalized moment of inertia $\bar{C}$, Love number $k_2$, and the initial rotation rate for each body where tides are acting.  Our implementation evaluates the tidal effects on the Earth-mass planet due to the gravitational forcing from each moon and the Sun.  

All our simulations assume the same value for the Love number ($k_2 = 0.298$).  We consider three dissipation scenarios for each moon system: $\tau = 0\, {\rm s}$ (no dissipation), $\tau = 100\, {\rm s}$ (weak tides), and $\tau = 698\, {\rm s}$ (Earth-Moon tidal dissipation). It is important to note that an Earth-mass exoplanet will not necessarily exhibit Earth-like tidal dissipation. The Earth's current dissipation rate is largely driven by bottom friction within its shallow ocean basins. An exoplanet lacking surface oceans, possessing a different interior rheology, or lacking similar topography, would likely exhibit weaker tidal friction. Therefore,  the $\tau = 0\, {\rm s}$ scenario allows us to account for these rheological differences and explore the dynamical stability of a more conservative system. 

We calculate the planet's moment of inertia as:

\begin{align}
I = \bar{C}_\oplus M_\oplus R_\oplus^{2} ,
\end{align}

\noindent where $\bar{C}_{\oplus} = 0.3308$ is the moment of inertia constant and $M_{\oplus}$ and $R_{\oplus}$ are the mass and radius of the planet.  We initialize the planet with a spin period of 6 hours ($P=6\, {\rm h}$) and an Earth-like obliquity of $\epsilon = 23.44^{\circ}$. We then calculate the spin magnitude as $\Omega_{0} = 2\pi/P$. The initial spin vector $\vec{\Omega}$ is defined in the $y$-$z$ plane as:

\begin{align}
{\vec\Omega} = (0, \Omega_0 \sin \epsilon, \Omega_0 \cos \epsilon),
\end{align}

\noindent where the system coordinates are rotated to align with the orbital angular momentum vector (the invariant plane).

Stellar tides account for the tidal forcing from the host star in the current day Earth-Moon analog system. While the Sun is significantly more massive than any moon, its large separation makes its tidal forcing comparatively weak. Because the tidal force scales as $M/d^{3}$, the ratio of the Earth's solar tide to its lunar tide is 

\begin{align}
\frac{F_{\odot}}{F_{\rm Luna}} \sim \frac{M_{\odot}/a^{3}} {M_{\rm Luna}/d^{3}} \approx 0.46,
\end{align}

\noindent where $M_{\odot}$ and $M_{\rm Luna}$ are the masses of the Sun and Moon, $a$ is the planet's semimajor axis, and $d$ is the planet to moon separation.

\begin{figure}[!htb]
\centering
\includegraphics[width=1\linewidth]{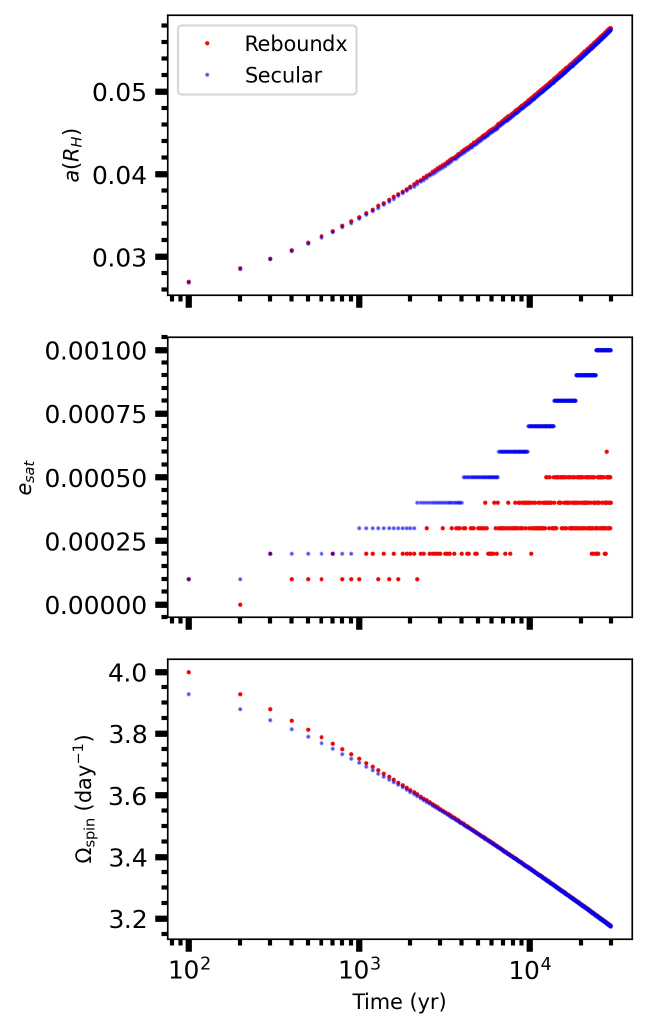}
\caption{Evolution of the semimajor axis $a$ in terms of the Hill radius ($R_{\rm H}$), the eccentricity $e_{\rm sat}$, and the spin rate of the planet $\Omega_{\rm spin}$ using the \texttt{REBOUNDx} and a secular tidal evolution model \citep{Barnes2017}.}
\label{Sec_vs_Reb_tides}
\end{figure}

To ensure that we implement the \texttt{tides\_spin} module correctly, we compare our \texttt{REBOUNDx} results to a secular tidal evolution model from \cite{Barnes2017}, which prescribes the semimajor axis, eccentricity, and spin rate evolution due to tides induced by external bodies.  In our test runs, we consider only the tidal evolution of an Earth-mass planet due to the combined tides from a moon and the Sun (i.e., $n=2$). Figure~\ref{Sec_vs_Reb_tides} compares the orbital evolution of the semimajor axis ($a$), the eccentricity ($e$), and the spin rate of the planet ($\Omega$) produced by \citet{Barnes2017} and \texttt{REBOUNDx} CTL model over $30,000$ yrs. Both models assume an Earth-Moon analog system with a single moon. We observe throughout all panels, both CTL models evolve similarly. The relations from \citeauthor{Barnes2017} are given by:

\begin{align}
\frac{da}{dt} 
  &= \frac{2a^2}{G M_1 M_2} 
     \sum_{i=1}^2 Z_{\rm i} \bigg(
        \cos(\psi_{\rm i}) 
        \frac{f_2(e)}{\beta^{12}(e)} 
        \frac{\Omega_{\rm i}}{n} 
        - \frac{f_1(e)}{\beta^{15}(e)} 
     \bigg)
\end{align}

\begin{multline}
\frac{de}{dt} = \frac{11ae}{2GM_1M_2} \sum_{i=1}^{2} Z_{\rm i} \left( \cos(\psi_{\rm i}) \frac{f_4(e)}{\beta^{10}(e)} \frac{\Omega_{\rm i}}{n} - \frac{18}{11} \frac{f_3(e)}{\beta^{13}(e)} \right)
\end{multline}

\begin{multline}
\frac{\mathrm{d}\Omega_{\rm i}}{\mathrm{d}t} = \frac{Z_{\rm i}}{2M_{\rm i} r_{\rm g}^2 R_{\rm i}^2 n} \left( 2\cos(\psi_{\rm i}) \frac{f_2(e)}{\beta^{12}(e)} - \right.\\
\left.\left[1 + \cos^2 \psi_{\rm i} \right]\frac{f_5(e)}{\beta^9(e)}\frac{\Omega_{\rm i}}{n}\right)
\end{multline}
where
\begin{align}
Z_{\rm i} 
  &\equiv 3 G^2 k_{2,{\rm i}} M_{\rm j}^2 (M_{\rm i} + M_{\rm j})
     \frac{R_{\rm i}^5}{a^9} \, \tau_{\rm i} \label{eq:Zi}
\end{align}
and
\begin{equation}
    \begin{aligned}
        \beta(e) &= \sqrt{1 - e^2}, \\
        f_1(e) &= 1 + \frac{31}{2} e^2 + \frac{255}{8} e^4 + \frac{185}{16} e^6 + \frac{25}{64} e^8, \\
        f_2(e) &= 1 + \frac{15}{2} e^2 + \frac{45}{8} e^4 + \frac{15}{16} e^6, \\
        f_3(e) &= 1 + \frac{4}{3} e^2 + \frac{8}{9} e^4 + \frac{5}{64} e^6, \\
        f_4(e) &= 1 + \frac{3}{2} e^2 + \frac{1}{8} e^4, \\
        f_5(e) &= 1 + 3e^2 + \frac{3}{8} e^4.
    \end{aligned}
\end{equation}

{In the CTL tidal model, the tidal time lag ($\tau$) controls the tidal dissipation rate and therefore the orbital evolution rate. As shown in Eqn. \eqref{eq:Zi} the  tidal forcing coefficient ($Z_i$) is proportional to ($\tau_i$), larger values of ($\tau$) generally produce faster tidal migration and eccentricity evolution.}

\begin{figure*}[!htb]
\centering
\includegraphics[width=1\linewidth]{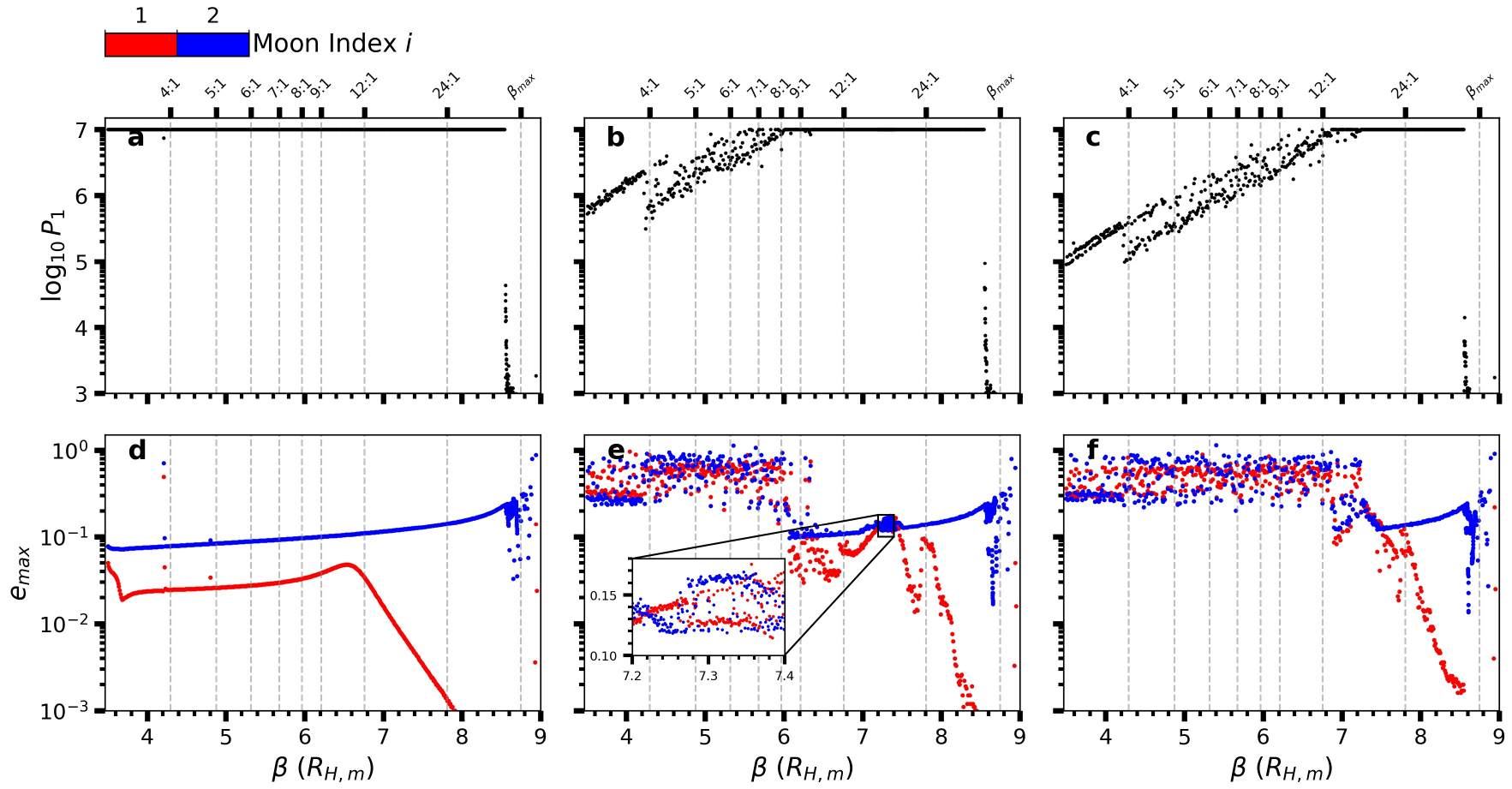}
\caption{The lifetime of the system with respect to the innermost orbit $P_{1}$ is plotted on a logarithmic scale for two Luna-mass moons. Alongside the maximum eccentricity reached by each moon is plotted (color-coded). The black-dashed line marks the $\beta_{\rm max}$ calculated earlier in the Eqn. \eqref{5}.{ The gray dashed lines and labels indicate the MMR locations predicted using Eqn. \eqref{MMR_Eqn}.}}
\label{fig:Luna_ecc_logt}  
\end{figure*}

\section{Results}  \label{sec: Results}

Using \texttt{REBOUND} and \texttt{REBOUNDx}, we investigate the dynamical stability of packed satellites around an Earth-mass planet that orbit a Sun-mass ($1\ M_{\odot}$) star. In these simulations, we track the orbital stability of two Luna-mass, three Pluto-mass, and five Ceres-mass moons for up to $10^{7}$ dynamical orbits of the innermost moon. We compare three tidal scenarios for each system: no tidal dissipation, weak tides, and Earth-like tidal dissipation. The planet begins on a mildly eccentric orbit ($e = 0.0167$), which introduces a forced eccentricity and perturbations from the host star.
\subsection{Luna Case Study} \label{Luna}

Figure~\ref{fig:Luna_ecc_logt} shows the results for two Luna-mass moons integrated up to $10^{7}$ orbits of the innermost moon, where $\log_{10}P_1$ corresponds to the lifetime of the system (Figures~\ref{fig:Luna_ecc_logt}a-\ref{fig:Luna_ecc_logt}c), and the maximum eccentricity (Figures~\ref{fig:Luna_ecc_logt}d-\ref{fig:Luna_ecc_logt}f) of each moon, both as functions of the spacing parameter $\beta$. The system lifetime for two Luna-mass moons without tides ($\tau = 0\, {\rm s}$; Figure~\ref{fig:Luna_ecc_logt}a) is insensitive to $\beta$ until $\beta_{\rm max}$ is reached.  For initial $\beta$ near the critical value of $\beta_{\rm max}$, the outer moon experiences strong perturbations from the Sun that propagate on short timescales and lead to scattering events.  Figure~\ref{fig:Luna_ecc_logt}d illustrates how the perturbations scale with increasing $\beta$ for the inner (red) and outer (blue) moon when tides are not implemented. For small values of $\beta$, the moon-moon interactions drive up the eccentricity of both moons to $e \sim 0.1$, where the moon-moon perturbations are non-negligible until $\beta \sim 6.5$.  Then the moon–moon interactions are largely decoupled where the outer moon's eccentricity is pumped by perturbations from the Sun.

For $\tau = 100\, {\rm s}$ (Figure~\ref{fig:Luna_ecc_logt}b), the system lifetime reaches the integration time for $6 \leq \beta \leq 8.6$. At lower spacing values ($\beta \leq 6$), low-order MMRs strongly excite the eccentricity (see Figure~\ref{fig:Luna_ecc_logt}e).  For $\beta \geq6$, the wider spacing weakens direct moon-moon interactions, so the maximum eccentricity remains low ($e \lesssim 0.1$), except near possible resonant locations. The spacing parameter $\beta$ determines the semimajor axis of subsequent moons, Eqn. \eqref{4}, and their orbital periods. As $\tau$ and $\beta$ increase, the moons encounter MMRs at different locations. We estimate resonance locations as a function of $\beta$, following \citet{Obertas2017}. Using Kepler's third law and the semimajor axis ratio $a_{i+1}/a_{i}$, we obtain

\begin{align} \label{MMR_Eqn}
    \beta\left(y\right) = \frac{1}{X}\,\frac{y^{2/3}-1}{y^{2/3}+1} ,
\end{align}

\noindent where $y=\frac{p_{i+1}}{p_{i}}$ is the period ratio of adjacent moons. As a result, we calculate that a high-order resonance falls within the range of $7 \leq \beta \leq 8$, which matches the structure seen in the zoomed-in region of Figure~\ref{fig:Luna_ecc_logt}e. {We check whether this is noise by increasing the sampling in $\beta$ to $0.001$ in the zoomed-in region. } We observe an eccentricity excitation in the outermost moon as it enters the MMR, which then decreases as the moon migrates outward. Supporting the MMR interpretation we see an anticorrelation between eccentricity of the two moons, where the inner moon's eccentricity is pumped due to the MMR. At larger $\beta$, stellar perturbations again excite the outer moon's eccentricity, leading to instability. 

Earth-like dissipation ($\tau = 698\, {\rm s}$; Figure~\ref{fig:Luna_ecc_logt}c) further restricts the range in beta, $7 \leq \beta \leq 8.7$, where the system lifetime is limited quasi-linearly for lower spacing values. Figure~\ref{fig:Luna_ecc_logt}f shows a similar excitation for lower values of $\beta$ because the stronger dissipation forces a faster outward migration for the inner moon and the increased spacing minimizes the subsequent number of MMR crossings. As tidal dissipation increases, the long-lived region is restricted, which is consistent with outward migration and resonance crossing that leads to collisions and/or scattering.  
 
Although \citet{Satyal2022} estimates that an Earth-mass planet could host $3 \pm 1$ Luna-mass satellites without tides, our numerical simulations show that, when tidal dissipation ($\tau$) is included, an Earth-mass planet hosts at most two Luna-mass satellites under certain restrictions. Some high-$\beta$ simulations survive the full integration, but the outer moon is strongly affected by stellar perturbations, so these systems are likely marginally stable.

\subsection{Pluto Case Study} \label{Pluto}

\begin{figure*}[!htb]
\centering
\includegraphics[width=1\linewidth]{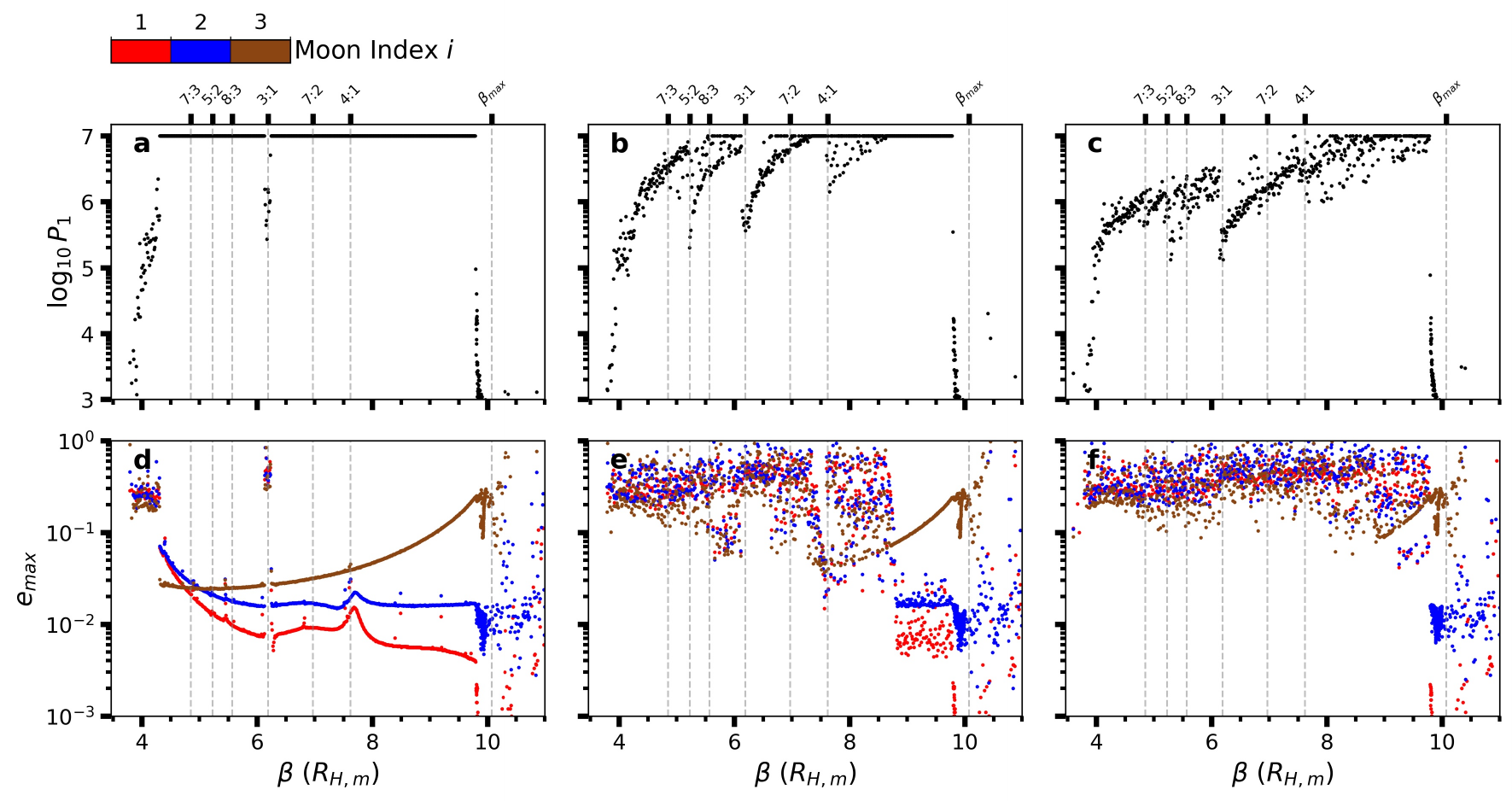}
\caption{Similar to Figure~\ref{fig:Luna_ecc_logt} but for 3 Pluto-mass satellites.}
\label{fig:Pluto_ecc_logt}  
\end{figure*}

We examine the evolution of three Pluto-mass satellite systems to potentially increase the number of potential moons.  The lower mass moons have a weaker interaction with each other, while migrating outward at a slower rate than when compared to Luna-mass moons.  

Figure~\ref{fig:Pluto_ecc_logt} follows a similar progression as with Figure~\ref{fig:Luna_ecc_logt}, where each panel shows how the system lifetime (Figures~\ref{fig:Pluto_ecc_logt}a-c) and maximum eccentricity of each moon (Figures~\ref{fig:Pluto_ecc_logt}d-f) varies with the assumed spacing parameter $\beta$. Figure~\ref{fig:Pluto_ecc_logt}a shows that the three-moon system with $\tau = 0\, {\rm s}$ is long-lived for $6 \leq \beta \leq \beta_{\rm max}$ with a small drop in lifetime around $\beta \approx 6.2$, which corresponds to the calculated location of a 3:1 MMR (see Eqn. \ref{MMR_Eqn}). The maximum eccentricity of each moon increases to $e \geq0.3$ near the resonance (Figure~\ref{fig:Pluto_ecc_logt}d). We additionally find a spike in eccentricity for the inner-moon pair (red and blue) around $\beta \approx 7.6$, {which coincides with the location of the 4:1 MMR predicted by Eqn. \eqref{MMR_Eqn}.} The outermost moon also experiences strong perturbations from the host star, resulting in a high eccentricity near $\beta_{\rm max}$. 

For $\tau = 100\, {\rm s}$ (Figure~\ref{fig:Pluto_ecc_logt}b), the moons are long-lived only in narrow ranges (e.g., $5.5 \leq\beta \leq6$ and $6.5 \leq\beta \leq9.5$). This is correlated to the low maximum eccentricity ($e \leq0.1$) seen in Figure~\ref{fig:Pluto_ecc_logt}e for these regions. Although the lifetimes are longer (up to the maximum simulation time) for $6.5 \leq \beta \leq 9.5$, the maximum eccentricity of the outermost moon continues to increase with increasing $\beta$, likely due to stronger stellar perturbations that eventually drive the outermost moon out of the system.

For Earth-like dissipation ($\tau = 698\, {\rm s}$; Figure~\ref{fig:Pluto_ecc_logt}c), the moons reach the maximum simulation time only for large spacing values $8\leq\beta \le9.7$. Although system lifetime is marginally obtained, the maximum eccentricity remains above $0.1$ (Figure~\ref{fig:Pluto_ecc_logt}f), which is likely due to strong moon-moon interactions alongside stellar and tidal perturbations. As a result, this induces a highly eccentric yet long-lived system, up to the system's maximum lifetime. However, a system beyond this lifetime will more than likely become dynamically unstable, where stellar perturbations will eject the outermost moon.

Our numerical simulations show that, when tidal dissipation ($\tau$) is included, an Earth-mass planet hosts at most three Pluto-mass satellites under certain restrictions. It then follows that a two Pluto-moon system appears the most likely outcome for either weak or strong tides.  This is in contrast to \citet{Satyal2022}, which estimated that an Earth-mass planet could host $4 \pm 1$ Pluto-mass satellites without tides. 

\subsection{Ceres Case Study} \label{Ceres}

\begin{figure*}[!htb]
\centering
\includegraphics[width=1\linewidth]{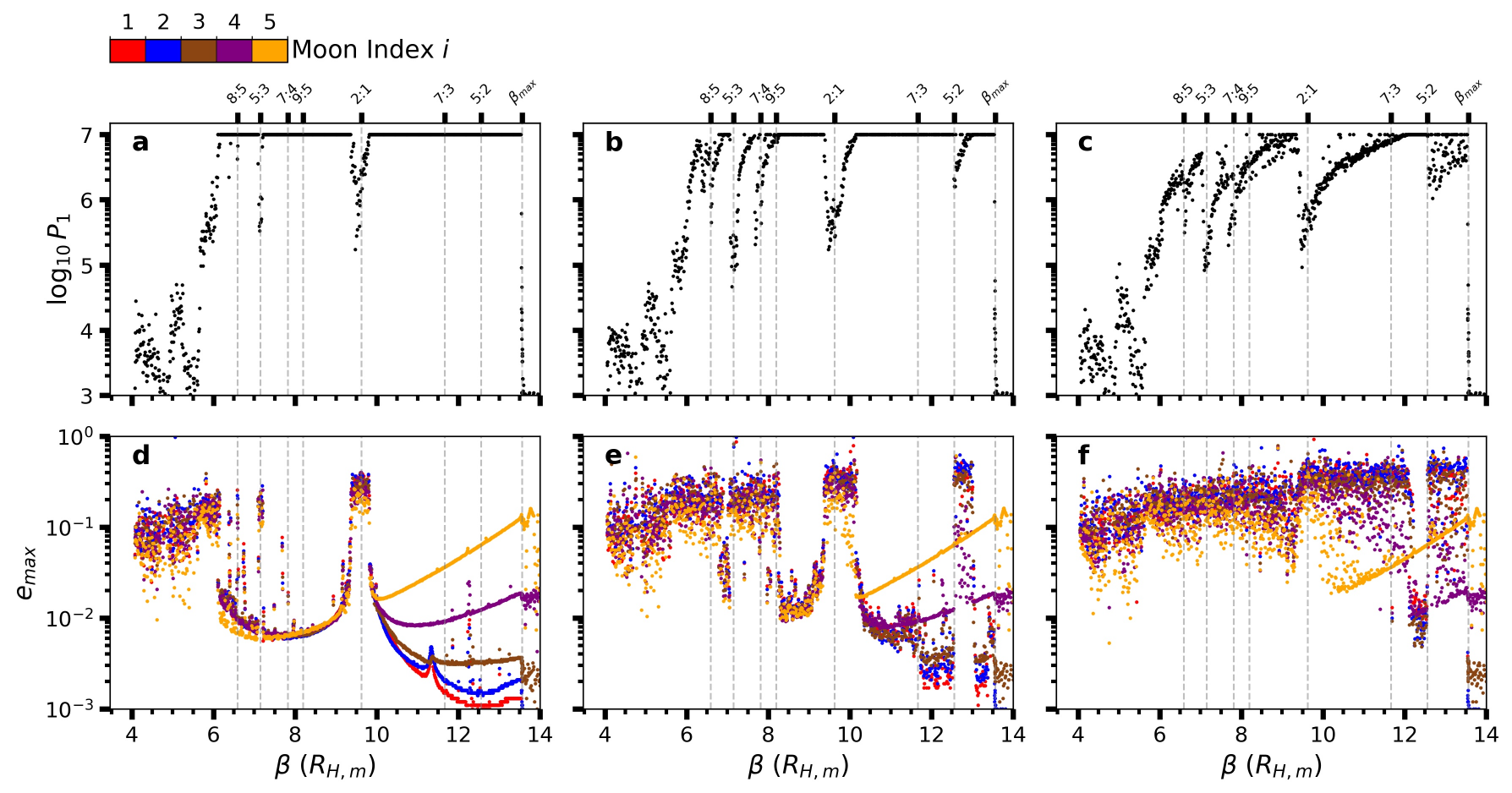}
\caption{Similar to Figure~\ref{fig:Luna_ecc_logt} but for 5 Ceres-mass satellites.}
\label{fig:Ceres_ecc_logt}  
\end{figure*}

We examine whether lower-mass satellites allow more moons to increase the system lifetime by modeling five Ceres-mass moons. Figure~\ref{fig:Ceres_ecc_logt}a shows that long-lived systems are possible for $6 \leq \beta \leq 13.5$, with narrow regions of instability. For $9.5 \leq \beta \leq 10$, we observe a drop in lifetime (Figure~\ref{fig:Ceres_ecc_logt}a) accompanied by a spike in eccentricity (Figure~\ref{fig:Ceres_ecc_logt}d), which is consistent with the location of the 2:1 MMR. We also note that the outermost moons experience strong perturbations from the host star, which excite their eccentricities for $\beta \geq10$.

Figures~\ref{fig:Ceres_ecc_logt}b and \ref{fig:Ceres_ecc_logt}e show our results for a weak tide ($\tau = 100\, {\rm s}$), which allows for narrower long-lived regions. The maximum eccentricity for these long-lived regions remains low ($e \leq 0.1$) similar to the tide-free scenario (Figure~\ref{fig:Ceres_ecc_logt}d) for the same values of $\beta$. For $\beta \geq 10$, the two outermost moons feel greater perturbations from the host star, which likely contribute to instability. In Figure~\ref{fig:Ceres_ecc_logt}e, we observe a notable spike in eccentricity for $12.5 \leq \beta \leq 13$, which corresponds to the 5:2 MMR.

Figures~\ref{fig:Ceres_ecc_logt}c and \ref{fig:Ceres_ecc_logt}f show the case of Earth-like dissipation. Although the moons are long-lived in some regions near $\beta \approx 12.2$, the maximum eccentricity for each moon continues to increase for $\beta \geq10$. Over the simulation the outermost moon orbits near the stability limit \citep{Rosario-Franco2020}, which increases its likelihood to exceed the stability limit at its apocenter. However, we observe a drop in eccentricity at $\beta \approx 12.5$, which  is anti-correlated with the feature seen in Figure~\ref{fig:Ceres_ecc_logt}e and may indicate an island of possible stability.

In the absence of tides, \citet{Satyal2022} showed that an Earth-mass planet could host $7 \pm 1$ Ceres-mass satellites.  Our numerical simulations show that, when tidal dissipation ($\tau$) is included, an Earth-mass planet hosts at most five Ceres-mass satellites under certain restrictions. We consider it more than likely that a $3-4$ Ceres-moon system appears stable for strong tides. 

\subsection{Time Series}

\begin{figure*}[!htb]
\centering
\includegraphics[width=1\linewidth]{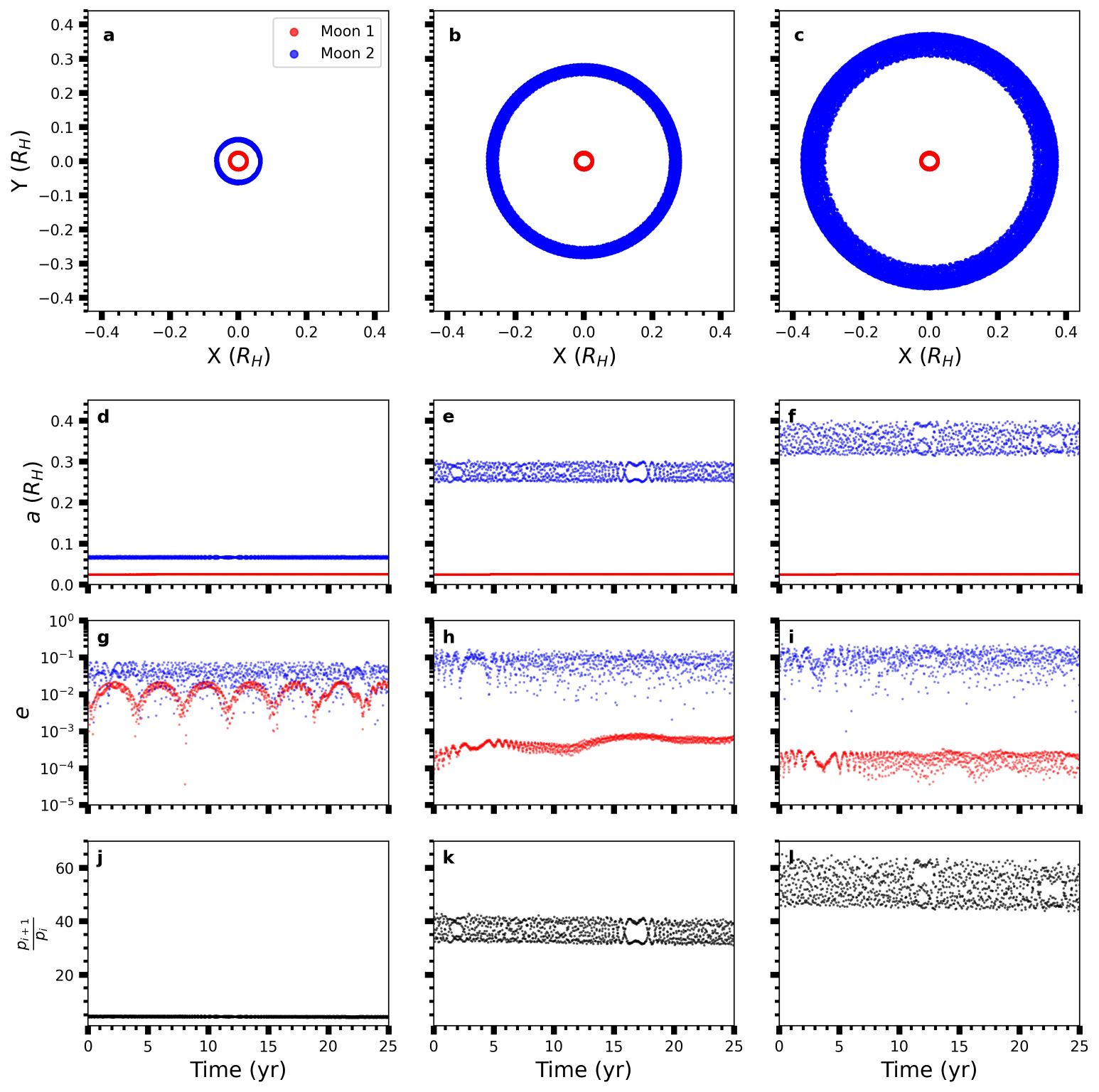}
\caption{The short-term orbital evolution of two Luna-mass moons ($\tau = 698\, {\rm s}$) separated by a factor $\beta$ relative to the planetary host, where the orbital distances are scaled relative to the Hill radius ($R_{\rm H}$).  Panels (a), (d), (g), (j) show the orbital evolution (starting with $\beta = 4.4$) with respect to the Cartesian coordinates, semimajor axis $a(R_{\rm H})$, eccentricity $e$, and the period ratio  $\frac{p_{i+1}}{p_{i}}$, respectively. Panels (b), (e), (h), and (k) show the evolution of the same parameters starting with $\beta = 8.2$, while panels (c), (f), (i), and (l) show the evolution for $\beta = 8.55$. }
\label{fig:Luna_time_series}
\end{figure*}

We analyze the orbital evolution of the moons under each moon-type assuming an Earth-like tidal dissipation factor ($\tau = 698\, {\rm s}$), where we track the semimajor axis and eccentricity of individual moons, and period ratios of moon pairs at an output frequency of $5\ P_{1}$. We investigate three stability regions defined by the spacing parameter ($\beta = 4.4$, $8.2$, and $8.55$) for a 25-year integration timescale. This timescale allows for thousands of orbits of the innermost moon, which is sufficient to identify immediate dynamical instabilities or chaotic behavior arising from the chosen $\beta$ value.  

For $\beta = 4.4$, the moons strongly interact with each other (Figure~\ref{fig:Luna_time_series}a), which causes their orbits to diffuse and increase the probability for the outer moon to escape over much longer timescales. Figure~\ref{fig:Luna_time_series}d shows that the semimajor axes of the inner and outer moons do not change significantly over the 25-year integration, although Figure~\ref{fig:Luna_time_series}g shows oscillations in eccentricity and indicates that the moons interact via an MMR (Figure~\ref{fig:Luna_time_series}j). The {inner moon's} eccentricity oscillates around $e \approx 0.01$. After ${\sim}$20~yr of evolution, the outer moon's eccentricity becomes proportional to the innermost moon (Figure~\ref{fig:Luna_time_series}g), which may indicate temporary capture into (or passage through) an MMR.  Because both moons remain well within the planetary Hill radius $R_{\rm H}$, the variability is driven primarily by mutual interactions rather than stellar perturbations at ($\beta = 4.4$)

As we increase the spacing parameter to $\beta = 8.2$, the outer moon can now interact more strongly with the host star than in the $\beta = 4.4$ spacing. (Figure~\ref{fig:Luna_time_series}b), although the faster outward migration of the inner moon still induces perturbations through high-order MMRs on longer timescales. The inner moon remains on a tight orbit (Figure~\ref{fig:Luna_time_series}e) on this timescale, where it executes roughly 1200 orbits within 25 yr. The period ratio evolves as the moons shift into and out of MMRs (Figure~\ref{fig:Luna_time_series}k). Around 15~yr, the outer moon appears to encounter a higher-order MMR, but it subsequently escapes due to continued outward tidal migration. The inner moon remains comparatively stable at $\beta = 8.2$, with $e \leq10^{-3}$, consistent with weaker moon-moon coupling at larger separations.

At even larger spacing ($\beta = 8.55$), the system approaches $\beta_{\rm max}$ and the outer moon evolves closer to the stability limit (Figure~\ref{fig:Luna_time_series}f and \ref{fig:Luna_time_series}i), where stellar perturbations become stronger. Once the outer moon's semimajor axis extends beyond the stability limit (${\sim}0.4R_{\rm H}$), the perturbations from the host star likely drive the outer moon to an instability, where it scatters off the host planet to escape or collides with the inner moon.


\begin{figure*}[!htb]
\centering
\includegraphics[width=1\linewidth]{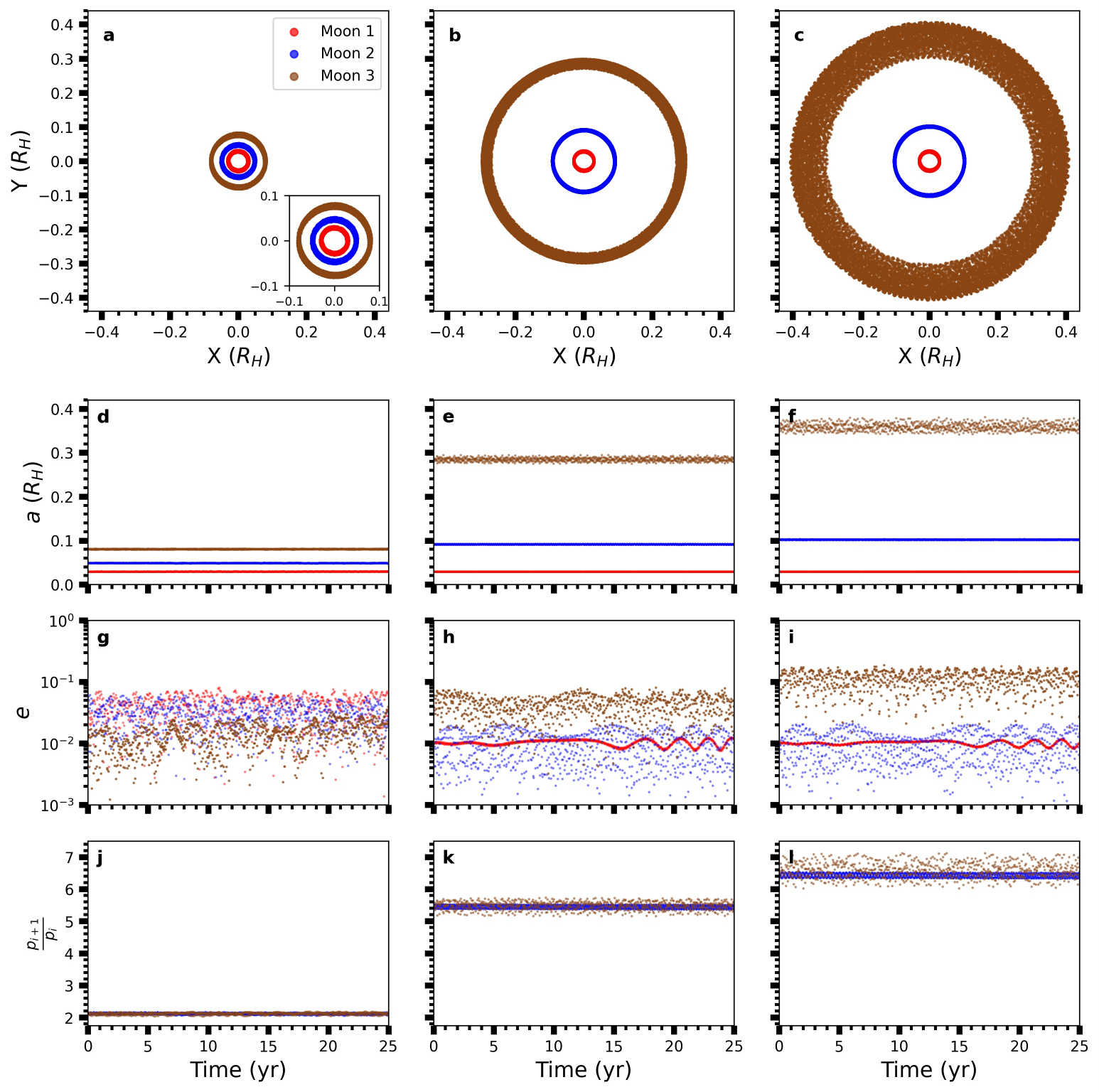}
\caption{Similar to Figure~\ref{fig:Luna_time_series} but for Pluto-mass moons at $\beta = 4.3$ (panel a), $\beta = 9.0$ (panel b), and $\beta = 9.7$ (panel c).}
\label{fig:Pluto_time_series}
\end{figure*}

As we increase the number of moons and decrease the mass to Pluto-mass moons, Figure~\ref{fig:Pluto_time_series} shows the orbital evolution of all three moons. The eccentricities of these moons remain low ($e \leq0.1$), but begin to grow as the system evolves  (Figure~\ref{fig:Pluto_time_series}g), where stellar perturbations begin to excite the eccentricity of the outermost moon. The eccentricity of the innermost adjacent moon pair remains higher than the outermost moon (Figure~\ref{fig:Pluto_time_series}g), more likely due to the close mutual Hill spacing, which allows strong moon-moon interactions.

As the mutual Hill radii increases to $\beta = 9$, the outermost moon's orbit expands outward, which allows for stronger perturbations from the Sun seen in Figure~\ref{fig:Pluto_time_series}h. With a larger $\beta$, this allows the eccentricity of the inner two moons to become anti-correlated (at $t\gtrsim 15\, {\rm yr})$, which indicates a likely interaction via MMRs. The period ratio between the outermost moon and the second moon begins to fluctuate around a 27:5 MMR due to resonance sweeping during outward tidal migration.

We evaluate the orbital elements near $\beta_{\rm max}$ for three Pluto-mass moons at $\beta = 9.7$. We see that the outer moon's eccentricity has increased to $e \geq0.1$ (Figure~\ref{fig:Pluto_time_series}i) alongside the expansion of the period ratio for the outermost adjacent pair (Figure~\ref{fig:Pluto_time_series}l). 

\begin{figure*}[!htb]
\centering
\includegraphics[width=1\linewidth]{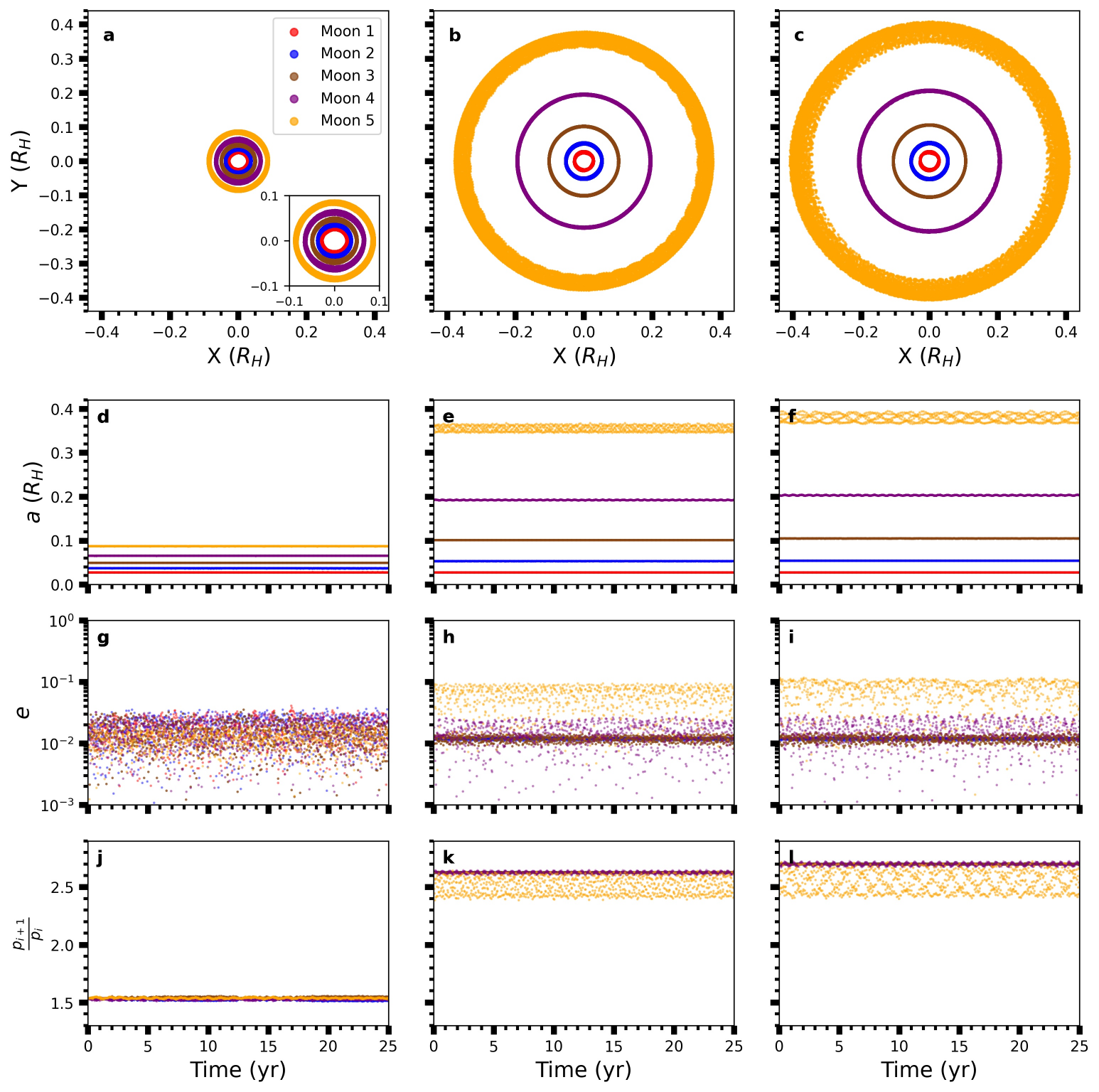}
\caption{Similar to Figure~\ref{fig:Luna_time_series} but for Ceres-mass moons at $\beta = 6.0$ (panel a), $\beta = 13.2$ (panel b), and $\beta = 13.55$ (panel c).}
\label{fig:Ceres_time_series}
\end{figure*}

We extend our analysis to five Ceres-mass moons (Figure~\ref{fig:Ceres_time_series}). The eccentricity of all moons fluctuates around $e \approx 0.01$ (Figure~\ref{fig:Ceres_time_series}g), indicating persistent dynamical coupling. The mutual perturbations remain strong due to a small enough spacing between neighboring moons. Figure~\ref{fig:Ceres_time_series}j supports this interpretation, as the period ratios overlap, indicating that the moons are likely not stable due to the multiple MMRs \citep{Wisdom1980,Mudryk2006}. Figures~\ref{fig:Ceres_time_series}e and \ref{fig:Ceres_time_series}f show coupled oscillations as the outermost moon experiences strong stellar perturbations in addition to interactions with its inner neighbor. This behavior is also reflected in the period ratios and eccentricity evolution (Figure~\ref{fig:Ceres_time_series}i and Figure~\ref{fig:Ceres_time_series}l), where the resonance structure becomes more apparent.

\subsection{Heat Maps}\label{sec:heat_maps}

\begin{figure*}[!htb]
\centering
\includegraphics[width=1\linewidth]{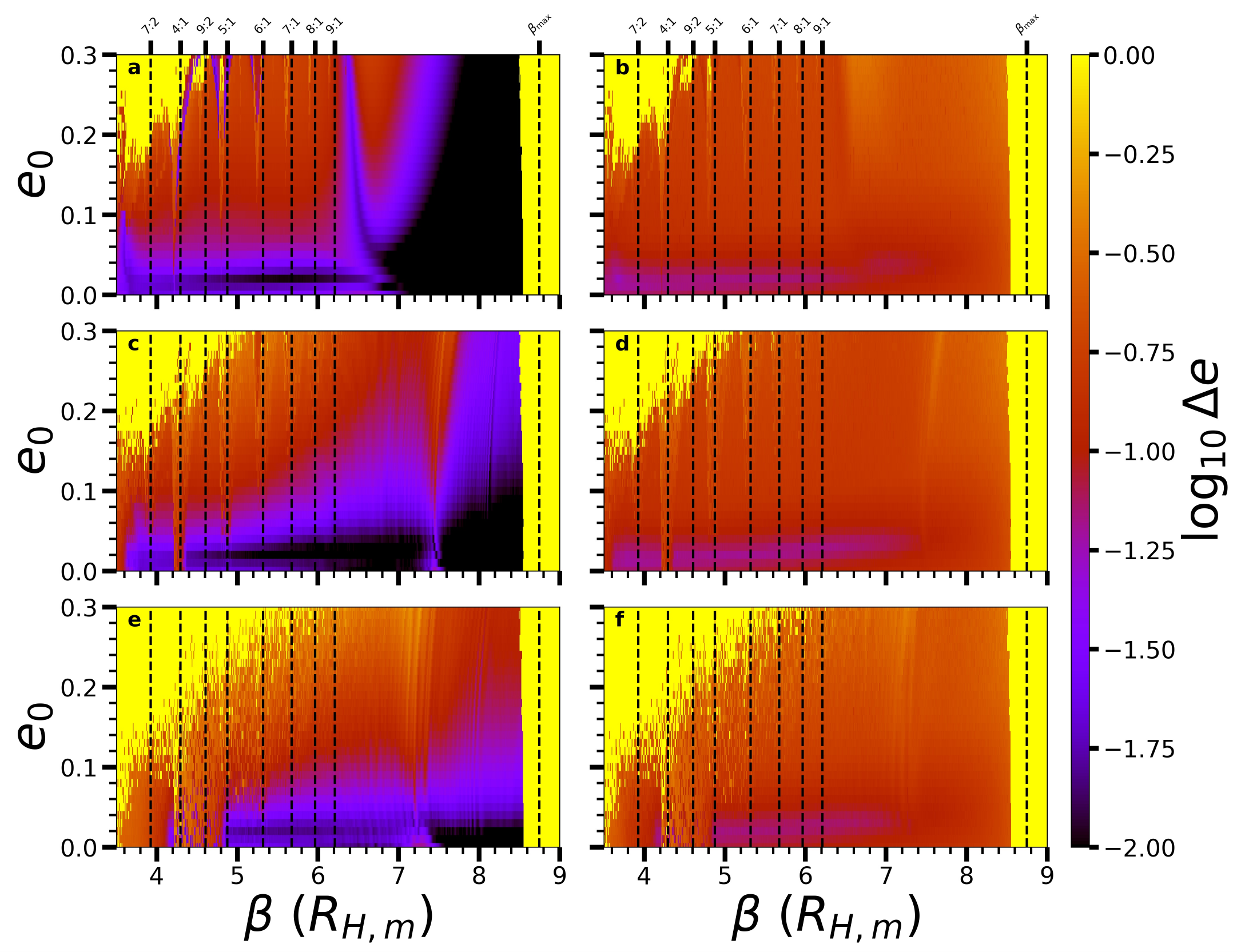}
\caption{Logarithmic difference between the maximum and minimum eccentricity ($\log_{10}\Delta e$) for a system with two Luna-mass moons simulated for $10^{5}$ orbits of the innermost moon $P_{1}$. These maps vary the initial eccentricity $e_{0}$ and the spacing parameter $\beta$. {Panels (a), (c), and (e) in the left column represent the results for the innermost moon when $\tau$ is $0\, {\rm s}$, $100\, {\rm s}$, and $698\, {\rm s}$, respectively while panels (b), (d), and (f) in the right column show the respective maps for the outermost moon.} The black-dashed lines denote the estimated locations of MMRs, as well as the $\beta_{\rm max}$.}
\label{fig:Luna_heat_map}
\end{figure*}

We explore the maximum variation in eccentricity ($\log_{10}\Delta e$) by tracking the maximum and minimum eccentricity during a simulation while varying the initial eccentricity ($e_{0}$) along with the spacing parameter $\beta$. We utilize this metric rather than traditional chaos indicators like MEGNO \citep{Cincotta2000,Cincotta2003} because our simulations introduce tidal dissipation where MEGNO is highly effective for conservative Hamiltonian systems. Since tidal dissipation actively drives secular orbital evolution, $\log_{10}\Delta e$ is more robust for tracking resonance bordering and orbital excitation that occurs as the moons migrate. The color-coded heatmaps (using $\log_{10}\Delta{e}$) allow us to visualize how elliptical an orbit can evolve over a simulation. The black regions ($\log_{10}\Delta e \approx -2.0$) indicate potentially dynamically stable regions because the system shows little eccentricity variation, implying weak orbital excitation and calm moon-moon interactions, while yellow regions ($\log_{10}\Delta e \approx 0.0$) indicate that the orbits become highly elliptical. In addition, the maps reveal fine resonant structures and help identify potentially chaotic regions.  The maximum spacing $\beta_{\rm max}$ should be interpreted as an idealized boundary in the absence of external forcing (e.g., tides or stellar perturbations), rather than a dynamical stability limit because these effects act over time to push the outermost moon's semimajor axis beyond the nominal stability limit. 

We limit the initial eccentricity up to $e_{0} = 0.3$, as higher values produce predominantly high eccentricity regions caused by the evolution of eccentricity \citep[][see their Figure 8]{Tamayo2021}. Figures~\ref{fig:Luna_heat_map}a, \ref{fig:Luna_heat_map}c, and \ref{fig:Luna_heat_map}e illustrate results for the innermost moon, while Figures~\ref{fig:Luna_heat_map}b, \ref{fig:Luna_heat_map}d, and \ref{fig:Luna_heat_map}f show the respective maps for the outermost moon. We assign unstable runs (as defined by the stopping conditions in Section~\ref{sec:meth}) a placeholder value of $e=1.5$ to mark unstable regions consistently.

For $\tau = 0\, {\rm s}$ (Figures~\ref{fig:Luna_heat_map}a and \ref{fig:Luna_heat_map}b), the maps resolve sharp, distinct MMRs for both moons. The innermost moon shows high eccentricity for ($2\sqrt3 \leq\beta \leq7$), while the outermost moon experiences large eccentricity variations, likely due to stellar perturbations. For $e_{0} \leq0.05$, the innermost remains stable across most values of $\beta$. For $e_{0} \geq0.05$, the innermost moon experiences eccentricity variations, consistent with the overlap of MMRs and/or evolution of the eccentricity \citep{Wisdom1980,Mudryk2006,Quillen2011,Lainey2020,Tamayo2021}. The innermost moon displays minimal eccentricity excitation in the dark region when $\beta \geq7$ up to $\beta_{\rm max}$, which we interpret as likely stable regions of the map.


Figures~\ref{fig:Luna_heat_map}c and \ref{fig:Luna_heat_map}d show the case with $\tau = 100\, {\rm s}$. As $e_{0}$ and $\beta$ increase, the innermost moon shows minimal eccentricity excitation until it enters a region near $\beta \approx 7.3$, which induces large eccentricity excitations. From $\beta = 4-5$, we find significant resonance broadening around the 4:1 and 5:1 MMRs. The U-shaped structure seen in Figure~\ref{fig:Luna_heat_map}a now appears to be blurred by outward tidal migration. Because the orbits drift rapidly under tidal forces, a moon that starts near a resonance can be dragged into it or swept across it. This causes the resonance region of influence to broaden overall. The outermost moon's resonance structures remain fairly similar, with only slight broadening due to its separation from the host planet and a weaker tidal dissipation. 

Figures~\ref{fig:Luna_heat_map}e and \ref{fig:Luna_heat_map}f show the scenario with Earth-like dissipation. With strong tidal dissipation, the system becomes highly eccentric and/or unstable at smaller values of $\beta$ and at larger $e_{0}$. In Figure~\ref{fig:Luna_heat_map}e, we resolve the MMR structure, however, the features now appear blended compared to the same regions in Figure~\ref{fig:Luna_heat_map}a. This indicates that the moons move in and out of resonances, thereby broadening the structure. The feature near $\beta \approx 7.1$ becomes more chaotic, with $\Delta e \geq0.1$. The outermost moon remains broadly similar, although a feature also forms near $\beta \approx 7.1$. Because it occurs at the same location for the innermost moon, we interpret this region as a MMR.

\begin{figure*}[!htb]
\centering
\includegraphics[width=1\linewidth]{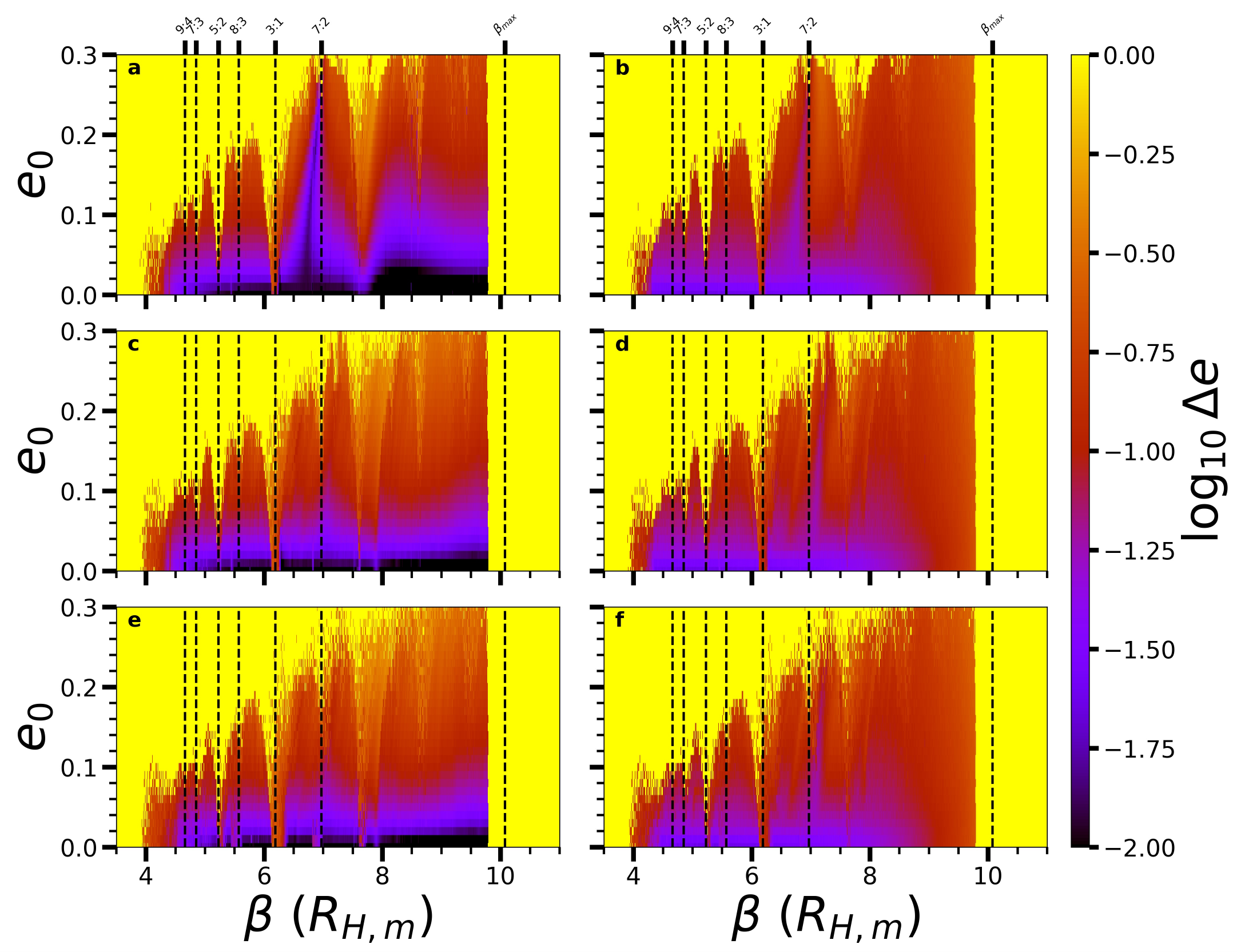}
\caption{{Similar to Figure~\ref{fig:Luna_heat_map} but with three Pluto-mass moons. The left column represents the innermost moon, while the right column represents the outermost moon.}}
\label{fig:Pluto_heat_map}
\end{figure*}

\begin{figure*}[!htb]
\centering
\includegraphics[width=1\linewidth]{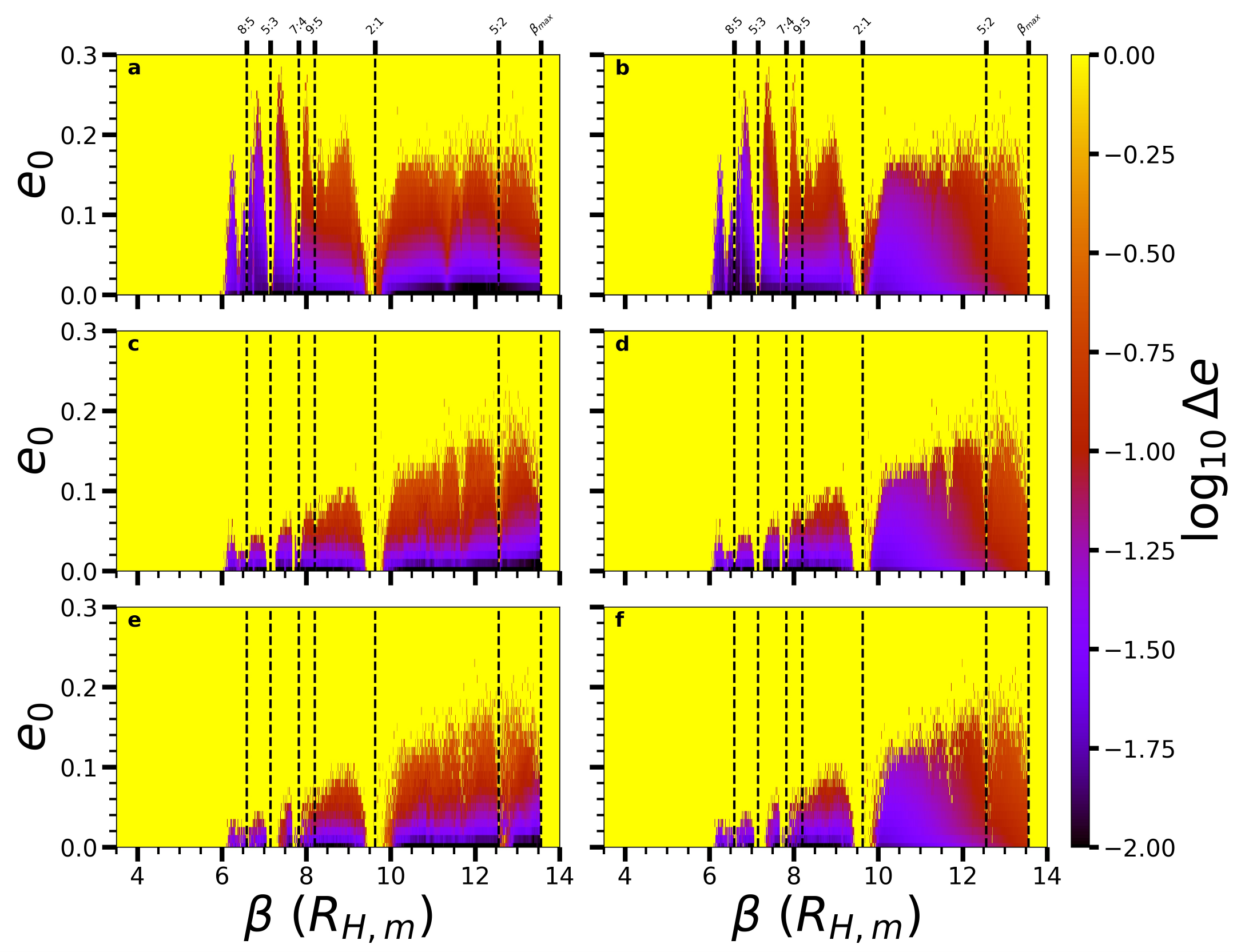}
\caption{{Similar to Figure~\ref{fig:Luna_heat_map} but with five Ceres-mass moons. The left column represents the innermost moon, while the right column represents the outermost moon.}}
\label{fig:Ceres_heat_map}
\end{figure*}

We extend the stability maps to three Pluto-mass moons (Figure~\ref{fig:Pluto_heat_map}) and five Ceres-mass moons (Figure~\ref{fig:Ceres_heat_map}). In both systems, increasing $e_{0}$ produces more chaotic behavior. For the Pluto-mass system, much of the parameter space became unstable under our stopping conditions, and stronger tidal dissipation broadened resonance-driven regions, where strong perturbations from MMRs or simply moon-moon interactions drive eccentricity excitations $(e\geq 0.1)$ across the parameter space (Figures~\ref{fig:Pluto_heat_map}e and \ref{fig:Pluto_heat_map}f).

The Ceres-mass system shows a less stable structure (Figure~\ref{fig:Ceres_heat_map}), with much smaller regions of low eccentricity excitation $\Delta e$ at smaller values of $e_{0}$. However, the maps still show large eccentricity variations across most of the parameter space.

\section{Conclusions} \label{sec: Conclusions}

We investigate the dynamical stability limits of an Earth-mass planet hosting multiple equal-mass moons with both gravitational and tidal forces. Our planet model includes Earth's radius, density, moment of inertia, obliquity, and Love number, where we assume a thin atmosphere exists that has a negligible tidal effect. While previous N-body simulations estimated that an Earth-mass planet could host up to several satellites \citep[e.g.,][]{Satyal2022}, our numerical simulations show that tidal forces drastically reduce the parameter space for dynamical stability. By utilizing \texttt{REBOUND} and \texttt{REBOUNDx}, we track the orbital evolution of Luna-, Pluto-, and Ceres-mass satellites over $10^7$ orbits of the innermost moon under multiple assumed tidal time lags for dissipation (no tidal dissipation, $\tau = 0\, {\rm s}$; weak tides, $\tau = 100\, {\rm s}$; and Earth-like tides, $\tau = 698\, {\rm s}$). {By varying ($\tau$), we explored how different tidal dissipation rates alter the orbital migration of packed moons. Larger tidal time lags accelerate tidal evolution, increasing the likelihood that moons encounter resonances and instability over time.}

Our results show that tidal forces act as a determinant for satellite architectures. An Earth-mass planet can host at most two Luna-mass satellites, which remain stable when $2\sqrt{3} \leq\beta \lesssim 8.7$ in the absence of tidal forces (Figure~\ref{fig:Luna_ecc_logt}a), whereas for weak tides the stability is shifted towards $6 \leq\beta \leq\beta_{\rm max}$ (Figure~\ref{fig:Luna_ecc_logt}b). Earth-like dissipation restricts the stability region to $7 \leq\beta \leq\beta_{\rm max}$ (Figure~\ref{fig:Luna_ecc_logt}c). We find that rapid outward tidal migration drives moons into MMRs, causing eccentric excitation that leads to orbital crossings (Figure~\ref{fig:Luna_time_series}) and eventual scattering events.

Systems with three Pluto-mass moons remain fairly stable under no and weak tidal forces. Although stable regions exist for $\tau = 0\, {\rm s}$ (Figure~\ref{fig:Pluto_ecc_logt}a) and $\tau = 100\, {\rm s}$ (Figure~\ref{fig:Pluto_ecc_logt}b), Earth-like dissipation restricts the system to a narrow region of $\beta$ (Figure~\ref{fig:Pluto_ecc_logt}c). Yet, at such a high spacing parameter, stellar perturbations are strongly felt as the outermost moon orbits near the upper stability limit.

Systems with five Ceres-mass moons show similar behavior to the systems with Pluto-mass moons but maintain regions of low eccentricity excitation near $\beta \approx 12$ and $\beta \approx 9$ under Earth-like dissipation (Figure~\ref{fig:Ceres_ecc_logt}c). However, the large spacing required for stability pushes the outermost moon near the stability limit, where stellar perturbations excite the eccentricity and likely destabilizes the system at {$t \geq 10^{7} P_{1}$, where $P_{1}$ is the innermost moon orbital period.}

\citet{2021MNRAS.500.1851K} introduced the ``exomoon corridor" that showed nearly half of all exomoons are expected to induce transit timing variation (TTV) signals within a few ($2-4$) epochs. \citet{Teachey2021} later extended this concept for multiple moon systems. These implications introduce new methods for detecting exomoon signals in transit, as well as other previously unidentified causes of TTV interference \citep{Kipping2023,Yahalomi2024}.{ Although an isolated exomoon candidate may display a clean transit signal, it may be difficult to distinguish the signal from other effects such as low-mass companion planets or stellar activity. Multi-moon systems preserve the ``exomoon corridor" phenomenon with TTVs appearing within the expected epoch. This suggests that packed moon systems may be at least as distinguishable, if not more so, than single-moon systems, while producing distinct TTV signals induced by moon-to-moon interactions.}  

The detection of exomoons around terrestrial planets remains challenging, with only a small number of current candidates identified through photometric methods \citep{Teachey2018,Kipping2022}. \citet{Pass2024} proposed to observe exoplanets TOI-700d and TOI-700e to search for Earth-Moon like analogs. Initially, the James Webb Space Telescope (JWST) was thought to be sensitive enough to detect these analogs. Even with JWST, our current observational capabilities remain insufficient to detect Earth-Moon analog systems \citep{Pass2026}. Despite observational limitations to terrestrial planets, \citet{Wilson2025} observed a free-floating planet (FFP; WISE 0855) to search for moon systems. {Their time-series observation revealed no exomoon transit, but their artificial transit modeling showed that JWST is sensitive to Galilean-mass moons around FFPs.} While these systems are not direct analogs to the Earth-Moon systems modeled here, they suggest that massive satellites around isolated objects may be detectable before the lower-mass, tightly packed moons around terrestrial planets.  A recent simulation framework based on relative astrometry of directly imaged giant exoplanets may improve the detectability of future exomoons \citep{Wagner2025}. In addition to individual star systems, theoretical studies have long indicated that exomoons may also exist around circumbinary planets, expanding our understanding of satellite architectures \citep{Quarles2012,Hamers2018}.

The proposed Habitable Worlds Observatory could also improve the prospects for detecting exomoons around terrestrial planets using a 6-8\,m space-based telescope that is designed to search for and image Earth-sized exoplanets in the habitable zones of their host stars \citep{NationalAcademiesofSciences2021,GoodisGordon2025,Ware2025}. In favorable systems, even Luna-mass satellites may be detectable \citep{Limbach2024}.  As future observations continue targeting habitable-zone exoplanets, our results imply that long-lived multi-moon systems around Earth-mass planets are possible, but strongly depend on tidal dissipation; if such conditions arise, tides affecting oceans likely must be weaker than those on Earth. Ultimately, if such systems exist, confirmation will rely on next-generation space-based telescopes (e.g., PLATO and HWO) to achieve the photometric sensitivity needed to isolate minuscule signals in the transit data. 

\section{Software and Data} \label{sec:cite}
The software and data used in this work are available at
\url{https://github.com/AlanB2006/Moon-Packing}
and are archived at\dataset[Moon Packing]{https://doi.org/10.5281/zenodo.20408204}.

\software{\texttt{REBOUND} \citep{Rein2012}, \texttt{REBOUNDx} \citep{Tamayo2020}}

\begin{acknowledgments}
This work used the Launch cluster at Texas A\&M High Performance Research Computing (HPRC) through allocation PHY240337 from the Advanced Cyberinfrastructure Coordination Ecosystem: Services \& Support (ACCESS) program, which is supported by U.S. National Science Foundation grants \#2138259, \#2138286, \#2138307, \#2137603, and \#2138296. This work was also supported by the Texas A\&M High Performance Research Computing (HPRC) facility through computing resources on the Launch cluster. The Launch cluster was funded by the National Science Foundation Campus Cyberinfrastructure (CC*) program under grant No. 2232895. 

A.B. and B.Q. gratefully acknowledge the College of Science and Engineering at East Texas A\&M University for their computational support of this work through the Regulus cluster in the Department of Physics and Astronomy. {The authors acknowledge the anonymous reviewer whose comments constructively improved various aspects of the manuscript.}

\end{acknowledgments}

\begin{contribution}

A.B. contributed to the early development of the simulation code, generated the figures and plots, and wrote the initial draft of the manuscript. B.Q. contributed substantially to the development of the code and revised the manuscript. M.R.-F. provided substantial feedback and revised portions of the manuscript.


\end{contribution}

%




\bibliography{references}
\bibliographystyle{mnras}



\end{document}